\documentclass[longauth]{aa} 
\usepackage{graphicx,natbib}
\bibpunct{(}{)}{;}{a}{}{,} 

\newcommand{\lapp}{\ensuremath{\stackrel{\scriptstyle <}{{}_{\sim}}}}
\begin{document}

\title{Contemporaneous observations of the radio galaxy NGC\,1275 from radio to very high energy $\gamma$-rays}

\author{
J.~Aleksi\'c\inst{1} \and
S.~Ansoldi\inst{2} \and
L.~A.~Antonelli\inst{3} \and
P.~Antoranz\inst{4} \and
A.~Babic\inst{5} \and
P.~Bangale\inst{6} \and
U.~Barres de Almeida\inst{6} \and
J.~A.~Barrio\inst{7} \and
J.~Becerra Gonz\'alez\inst{8} \and
W.~Bednarek\inst{9}
% ...
(\textit{The MAGIC Collaboration}),\\
 and B.~Balmaverde\inst{31} \and
 J.~Kataoka\inst{32} \and
 R.~Rekola\inst{20} \and
 Y.~Takahashi\inst{32}
}
\institute { IFAE, Edifici Cn., Campus UAB, E-08193 Bellaterra, Spain
\and Universit\`a di Udine, and INFN Trieste, I-33100 Udine, Italy
\and INAF National Institute for Astrophysics, I-00136 Rome, Italy
\and Universit\`a  di Siena, and INFN Pisa, I-53100 Siena, Italy
\and Croatian MAGIC Consortium, Rudjer Boskovic Institute, University of Rijeka and University of Split, HR-10000 Zagreb, Croatia
\and Max-Planck-Institut f\"ur Physik, D-80805 M\"unchen, Germany
\and Universidad Complutense, E-28040 Madrid, Spain
\and Inst. de Astrof\'isica de Canarias, E-38200 La Laguna, Tenerife, Spain
\and University of Lodz, PL-90236 Lodz, Poland
\and Deutsches Elektronen-Synchrotron (DESY), D-15738 Zeuthen, Germany
\and ETH Zurich, CH-8093 Zurich, Switzerland
\and Universit\"at W\"urzburg, D-97074 W\"urzburg, Germany
\and Centro de Investigaciones Energ\'eticas, Medioambientales y Tecnol\'ogicas, E-28040 Madrid, Spain
\and Technische Universit\"at Dortmund, D-44221 Dortmund, Germany
\and Inst. de Astrof\'isica de Andaluc\'ia (CSIC), E-18080 Granada, Spain
\and Universit\`a di Padova and INFN, I-35131 Padova, Italy
\and Universit\`a dell'Insubria, Como, I-22100 Como, Italy
\and Unitat de F\'isica de les Radiacions, Departament de F\'isica, and CERES-IEEC, Universitat Aut\`onoma de Barcelona, E-08193 Bellaterra, Spain
\and Institut de Ci\`encies de l'Espai (IEEC-CSIC), E-08193 Bellaterra, Spain
\and Finnish MAGIC Consortium, Tuorla Observatory, University of Turku and Department of Physics, University of Oulu, Finland
\and Japanese MAGIC Consortium, Division of Physics and Astronomy, Kyoto University, Japan
\and Inst. for Nucl. Research and Nucl. Energy, BG-1784 Sofia, Bulgaria
\and Universitat de Barcelona (ICC/IEEC), E-08028 Barcelona, Spain
\and Universit\`a di Pisa, and INFN Pisa, I-56126 Pisa, Italy
\and now at Ecole polytechnique f\'ed\'erale de Lausanne (EPFL), Lausanne, Switzerland
\and now at Department of Physics \& Astronomy, UC Riverside, CA 92521, USA
\and now at Finnish Centre for Astronomy with ESO (FINCA), Turku, Finland
\and also at Instituto de Fisica Teorica, UAM/CSIC, E-28049 Madrid, Spain
\and Now at Stockholms universitet, Oskar Klein Centre for Cosmoparticle Physics
\and now at GRAPPA Institute, University of Amsterdam, 1098XH Amsterdam, Netherlands
\and INAF - Osservatorio Astrofisico di Torino, Strada Osservatorio 20, 10025 Pino Torinese, Italy
\and Waseda University, Tokyo, Japan
}
\date{Received: October~28, 2013 / Accepted: January~23, 2014}

\offprints{Pierre Colin (colin@mppmu.mpg.de), Saverio Lombardi (saverio.lombardi@oa-roma.inaf.it), Fabrizio Tavecchio (fabrizio.tavecchio@brera.inaf.it), Doroth\'ee Hildebrand (dorothee.hildebrand@phys.ethz.ch), Elina Lindfors (elilin@utu.fi)}


  \abstract
   {}
   {The radio galaxy \object{NGC\,1275}, recently identified as a very high energy (VHE, $>$100\,GeV) $\gamma$-ray emitter by MAGIC,
    is one of the few non-blazar active galactic nuclei detected in the VHE regime.
    The purpose of this work is to better understand the origin of the $\gamma$-ray emission and locate it within the galaxy.
    }
   {We studied contemporaneous multi-frequency observations of \object{NGC\,1275} and modeled the overall spectral energy distribution.
    We analyzed unpublished MAGIC observations carried out between October 2009 and February 2010, and the previously published observations taken between August 2010 and February 2011.
    We studied the multi-band variability and correlations by analyzing data of \textit{Fermi}-LAT in the 100\,MeV--100\,GeV energy band, as well as \textit{Chandra} (X-ray), KVA (optical), and MOJAVE (radio) data
    taken during the same period.
    }
  {Using customized Monte Carlo simulations corresponding to early MAGIC stereoscopic data, we detect \object{NGC\,1275} also in the earlier MAGIC campaign.
  The flux level and energy spectra are similar to the results of the second campaign. The monthly light curve above 100\,GeV shows
  a hint of variability at the 3.6\,$\sigma$ level. In the \textit{Fermi}-LAT band, both flux and spectral shape variabilities are reported.
  The optical light curve is also variable and shows a clear correlation with the $\gamma$-ray flux above 100\,MeV.
  In radio, three compact components are resolved in the innermost part of the jet. One of these components shows a similar trend
  as the {\it Fermi}-LAT and KVA light curves.
  The $\gamma$-ray spectra measured simultaneously with MAGIC and \textit{Fermi-}LAT from 100\,MeV to 650\,GeV
  can be well fitted either by a log-parabola or by a power-law with a sub-exponential cutoff for the two observation campaigns.
  A single-zone synchrotron-self-Compton model, with an electron spectrum following a power-law with an exponential cutoff,
  can explain the broadband spectral energy distribution and the multi-frequency behavior of the source.
  However, this model suggests an untypical low bulk-Lorentz factor or a velocity alignment closer to the line of sight than the pc-scale radio jet.
}
   {}

\keywords{galaxies: active - galaxies: jets - galaxies: individual (NGC\,1275/3C\,84) - gamma rays: galaxies }

\titlerunning{NGC\,1275 multi-frequency observation}

\authorrunning{Aleksi\'c et~al. (MAGIC collaboration)}

\maketitle


\section{Introduction}
\label{sec:1}

The radio galaxy \object{NGC\,1275} is the central dominant galaxy of the Perseus cluster (\object{Abell 426}).
It is a well-known active galactic nucleus (AGN) already included in the original study of \cite{sey43}.
In fact, it is a rather complex object with very peculiar characteristics that might be the result of massive gas accretion from the cluster cooling flow or from a recent merging event \citep{con01}. The main features, in addition to the central AGN, are remarkable gas filaments that are tens of kpc long and emit strong hydrogen lines \citep{lyn70}, and a foreground galaxy that falls toward NGC\,1275 \citep{cau92}.

In radio, the emission is dominated by a very bright compact source (\object{3C\,84}) at the center of the galaxy.
The core-dominated morphology with asymmetrical jets at both kpc \citep{ped90} and pc scales \citep{asa06} looks like a Fanaroff-Riley type I (FR~I) radio galaxy with a jet axis relatively close to the line of sight. The faint counter jet measured with Very Long Baseline Interferometry (VLBI) suggests a jet angle of 30$^\circ$--55$^\circ$ in the core region \citep{ver94,wal94,asa06}. At larger scales, there is evidence of jet bending \citep{ped90}.
The flux and morphology of the core radio emission are variable. Recent observations showed that an outburst is on-going since 2005 \citep{nag10}.
At sub-pc scales, a new component appeared near the nucleus in 2007 and continues to grow in flux as it travels away from the nucleus. Since 2010, its brightness at 15\,GHz has exceeded that of the nucleus \citep{nag12}.

The X-ray emission is dominated by the thermal emission of the intracluster medium cooling flow in the cluster central region \citep{chu03,fab11}. However, a non-thermal component from the nucleus is detected with high-resolution instruments \citep{chu03,bal06}, which follows a hard power-law spectrum (photon index $\Gamma$=1.6$\pm$0.1) in the 0.5--10\,keV range. Non-thermal emission also appears in hard X-rays above $\sim$20\,keV \citep{aje09,eck09}, which matches the extrapolation of the AGN power-law X-ray emission well.

In $\gamma$-rays, emission from both the cluster and the AGN was expected.
A first hint of high-energy emission from the NGC\,1275 region was found in the COS~B data [1975--1982] \citep{str83}
but was not confirmed by the next instrument \textit{CGRO}-EGRET, which started observations one decade later [1991--2000] \citep{rei03}.
After the first four months of the all-sky survey with \textit{Fermi}-LAT in 2008, $\gamma$-ray emission from NGC\,1275 was clearly established \citep{abd09}. The measured flux above 100\,MeV was seven times higher than the \textit{CGRO}-EGRET upper limit, suggesting strong variability of the source. Subsequent \textit{Fermi}-LAT observations confirmed this variability, revealing variation time-scales as rapid as a week \citep{kat10,bro11}.

At very high energies (VHE, $>$100\,GeV), NGC\,1275 has been observed without success by many Cherenkov telescope experiments: HEGRA~\citep{goe01}, Whipple-10\,m~\citep{per06}, VERITAS~\citep{acc09}, and MAGIC-I~\citep{ale10b}.
The first detection was recently reported by MAGIC in stereoscopic mode \citep{ale12a}. The 70--500\,GeV energy spectrum is much steeper than the 0.1--20\,GeV spectrum measured by \textit{Fermi}-LAT, suggesting a break or a cutoff at a few tens of GeV.
The angular resolution of the $\gamma$-ray telescopes is not sufficient to determine the origin of the emission within the galaxy, but the rapid
variability seen by \textit{Fermi}-LAT implies a very compact emission region, most likely from the inner part of the AGN.
Additionally, the expected cluster $\gamma$-ray emission
probably extends across a region of several hundred kpc \citep{Pin10}, and then must be constant on human time-scales.
This emission may be detectable when the AGN is quiet \citep{col10} or above the high-energy cutoff of the AGN \citep{ale12b}.


In this work, we study contemporaneous multi-frequency observations of the AGN emission.
We analyzed the MAGIC data taken in stereoscopic mode from October 2009 to February 2011 as well as \textit{Fermi}-LAT data, X-ray \textit{Chandra}
data, optical data of KVA and NOT, and radio data of the MOJAVE program at VLBA from the same observation period. 
Together with additional information from archival data, we discuss a possible emission scenario and try to model the source as a misaligned BL Lac object.


\section{Observations and data analysis}
\label{sec:2}


\subsection{MAGIC}
\label{subsec:2.1}

MAGIC is a system of two imaging atmospheric Cherenkov telescopes 17\,m in diameter
located at the Roque de los Muchachos observatory ($28.8^\circ$N, $17.8^\circ$W, 2200\,m a.s.l.),
on the Canary Island of La Palma. The MAGIC telescopes have been operating in stereoscopic mode since October 2009
with a sensitivity $\leq$0.8\% of the Crab Nebula flux, for energies above $\sim$300\,GeV,
in $50$\,h of observations \citep{ale12c}. The trigger system, optimized for the lowest energies, enables one
to study $\gamma$-ray emission down to $50$\,GeV.

The MAGIC telescopes observed NGC\,1275 during two distinct observational campaigns performed between October 2009
and February 2010 ($\sim$45.3\,h), and August 2010 and February 2011 ($\sim$53.6\,h).
Both campaigns were carried out in stereoscopic mode, but under different global trigger configurations.

The first campaign (Camp.\,1) occurred partially during the commissioning phase of the second telescope (MAGIC-II).
It was performed in the so-called soft-stereo trigger mode, with the first telescope (MAGIC-I) trigger working in single mode
and the second telescope (MAGIC-II) recording only events triggered by both telescopes.
This campaign resulted in the discovery of $\gamma$-ray emission above $300$\,GeV from \object{IC\,310}, another radio galaxy
of the Perseus cluster \citep{ale10a,ale13}.

In the second campaign (Camp.\,2), observations were instead taken in the standard full-stereo trigger mode 
(recorded events are triggered simultaneously by both telescopes). It resulted in the first detection
of NGC\,1275 at energies above $100$\,GeV \citep{ale12a}.

The soft-stereo trigger allows us to analyze MAGIC-I data in single mode but has a slightly
higher energy threshold for stereoscopic data than the full-stereo trigger. However, above $\sim$150\,GeV
data collected in both campaigns are almost identical and can be treated as a uniform data sample.
This was the case for the study of the $\gamma$-ray emission induced by cosmic-ray population in the Perseus cluster~\citep{ale12b}.
Instead, for data analysis below 150\,GeV (as performed here), the two campaigns must be analyzed separately.

%

Both surveys were performed in the false-source tracking (wobble) mode \citep{fom94}, 
with data equally split into two (Camp.\,1) or four (Camp.\,2) pointing positions located 
symmetrically at $0.4^\circ$ from NGC\,1275. Observations were carried out at low zenith angles ($<$35$^\circ$),
which resulted in an analysis energy threshold (defined as the peak of the $\gamma$-ray energy distribution
for a Crab-like spectrum) of 100\,GeV.

The data analysis was performed using the standard software package MARS~\citep{ali09},
which includes the latest standard routines for stereoscopic analysis~\citep{ale12c,lom11}.
The $\gamma$-ray selection cuts were optimized by means of contemporaneous Crab Nebula data and Monte Carlo simulations.
The background was estimated from mirror regions (off-regions) corresponding to the source position in the camera during the other wobble pointing.
Thus, depending on the number of wobble positions used in the campaign, a single off-region (Camp.\,1) or three off-regions (Camp.\,2) were considered.
While the standard Monte Carlo simulations were used for the analysis of Camp.\,2 data \citep[reported in][]{ale12a}, the data analysis of the first campaign required new dedicated simulations to fully take into account the non-standard trigger condition at the lowest energies ($<$150\,GeV).

After applying standard quality checks based on the rate of the stereo events and the 
distributions of basic image parameters, $39.0$\,h (Camp.\,1) and $45.7$\,h (Camp.\,2)
of data were selected to derive the results presented here. The rejected data were affected mainly 
by poor atmospheric conditions during the data taking.

\begin{figure}
\centering
\includegraphics[width=0.42\textwidth]{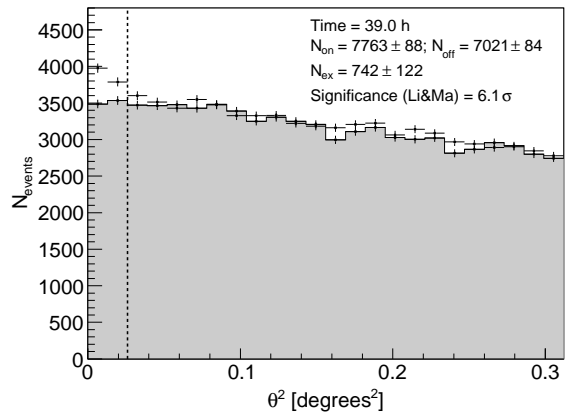}
\caption
{
$\theta^{2}$ distributions of the NGC\,1275 signal and background estimation from $39.0$\,h 
of MAGIC stereo observations taken between October 2009 and February 2010, in soft-trigger
stereo mode (see~\ref{subsec:2.1}) with an energy threshold of 100\,GeV. The region between 
zero and the vertical dashed line (at 0.026\,$\mbox{degrees}^{2}$) represents the signal region.
}
\label{fig:theta2}
\end{figure}

The $\gamma$-ray signal from the NGC\,1275 is detected only at low energies.
Fig.~\ref{fig:theta2} shows the $\theta^{2}$ plot\footnote{The parameter $\theta^{2}$ is the square angular
distance between the reconstructed source of the events and the nominal positions of the expected source.}
obtained with Camp.\,1 data (October~2009 -- February~2010) for analysis cuts corresponding to an energy threshold of 100\,GeV.
We found an excess of 742$\pm$122 events in the fiducial signal region with $\theta^{2} < 0.026$\,$\mbox{degree}^{2}$,
corresponding to a significance of 6.1\,$\sigma$, calculated according to the Eq.~17 of~\cite{lima83}.
During the second campaign (August~2010 -- February~2011), an excess of 522$\pm$81 events above the same energy threshold was detected
with a significance of 6.6\,$\sigma$. Detailed Camp.\,2 results are reported in \cite{ale12a}.

\begin{figure}
\centering
\includegraphics[width=0.47\textwidth]{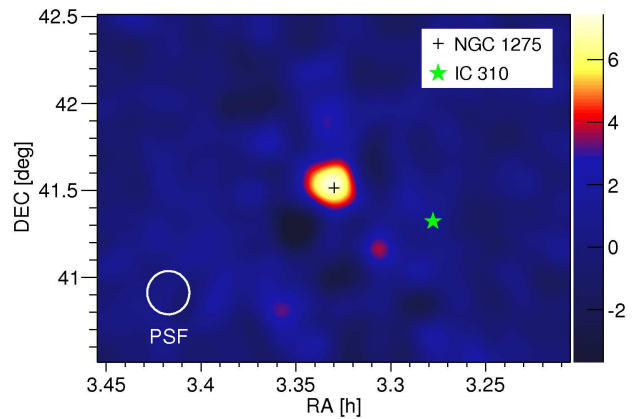}
\caption
{
Significance skymap (J2000) of the NGC\,1275 region from $39.0$\,h of MAGIC stereo observations taken 
between October 2009 and February 2010, with a low energy threshold of $100$\,GeV.
The point-spread function (PSF) of about 0.12$^\circ$ is also displayed.
}
\label{fig:skymap}
\end{figure}

The significance skymap above 100\,GeV for the first campaign is shown in Fig.~\ref{fig:skymap}.
The central hot spot ($>$6\,$\sigma$) corresponding to the \object{NGC\,1275} position is consistent with a point-like source.
Conversely to the skymap above 400\,GeV reported in \citet{ale10a} for the same period,
the radio galaxy \object{IC\,310} is not visible in the $\sim$100\,GeV skymap due to its very hard spectrum.

\begin{figure}
\centering
\includegraphics[width=0.45\textwidth]{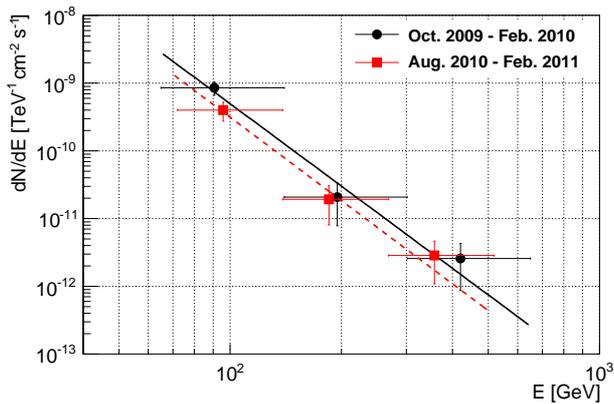}
\caption{NGC\,1275 differential energy spectrum measured with MAGIC during two observation campaigns and their associated power-law fits.
Camp.\,1 data are in black with a solid line fit and Camp.\,2 data are in red with a dashed line fit.}
\label{fig:spectrum}
\end{figure}

The unfolded differential energy spectra of the source derived from the two observational campaigns are shown in Fig.~\ref{fig:spectrum}.
The error bars represent the statistical uncertainties only, but for the power-law fit parameterization the systematic effects\footnote{The systematic errors of the flux normalization and the energy spectral slope considered here have been estimated to be lower than $23\%$  and $\pm0.3$, respectively, whereas the systematic uncertainty on the energy scale is $17\%$. These values are more conservative than those presented in~\cite{ale12c}, because of the flux weakness and the spectral steepness of NGC\,1275, as measured by MAGIC.} are also taken into account.
For Camp.\,1, the spectrum between 65\,GeV and 650\,GeV can be described by a simple power-law ($\chi^{2}/n_{dof}=4.86/3$)
\begin{equation}
\frac{\mbox{d}F}{\mbox{d}E} = f_{0} \left(\frac{E}{\mathrm{100\,GeV}}\right)^{-\Gamma},
\end{equation}
with a photon index $\Gamma=4.0\pm0.5_{stat}\pm0.3_{syst}$
and a normalization constant at 100\,GeV $f_{0} =(5.0 \pm 0.9_{stat} \pm 1.1_{syst}) \times 10^{-10} \mathrm{cm^{-2} s^{-1} TeV^{-1}}$.
The spectrum of Camp.\,2 corresponds to the result reported in \cite{ale12a}: $f_{0} =(3.1 \pm 1.0_{stat} \pm 0.7_{syst}) \times 10^{-10} \mathrm{cm^{-2} s^{-1} TeV^{-1}}$ and  $\Gamma=4.1\pm0.7_{stat}\pm0.3_{syst}$.
The energy range of the Camp.\,1 spectrum is slightly larger than that of Camp.\,2.
This arises because the Camp.\,1 spectrum must have slightly larger bin widths in energy to fulfill the significance requirement of each bin.

The mean flux above 100\,GeV during Camp.\,1, $(1.6\pm0.3_{stat}$~$\pm$~$0.3_{syst})\times10^{-11}~\mathrm{cm^{-2}~s^{-1}}$,
corresponds to about 3$\%$ of the Crab nebula flux.
It is slightly higher than the Camp.\,2 results, $(1.3\pm0.2_{stat}$~$\pm$~$0.3_{syst})\times10^{-11}~\mathrm{cm^{-2}~s^{-1}}$, but both campaigns agree within the uncertainties. No indication of spectral index variability has been found between the two campaigns.
The monthly light curve (LC) of NGC\,1275 above 100\,GeV is shown in the top panel of Fig.~\ref{fig:MWL_LC}.
While the Camp.\,2 LC is quite consistent with a constant flux ($\chi^{2}/n_{dof}=7.4/4$, probability = 0.29), a hint of variability can be derived for the first campaign. Indeed, fitting the Camp.\,1 LC with a constant flux yields a $\chi^{2}/n_{dof}=22.9/4$ (probability = 1.3$\times$10$^{-4}$), corresponding to a significance for a monthly variable emission of 3.6\,$\sigma$.

\subsection{\textit{Fermi}-LAT}
\label{subsec:2.2}

The Large Area Telescope (LAT) is a pair-conversion telescope onboard the \textit{Fermi $\gamma$-ray Space Telescope}, designed to cover the energy band from 20\,MeV to higher than 300\,GeV \citep{atw09}. The observations used here comprise all scientific data obtained between August 4, 2008  and February 21, 2011. We applied a zenith angle cut of 100$^\circ$ to greatly reduce $\gamma$-rays from Earth's limb. The same zenith cut is also accounted for in the exposure calculation using the LAT Science Tool gtltcube. We used the ``Source'' class events \citep{ack12},
which is the recommended class for regular analysis.
In our analysis, Science Tools version v9r27p1 and instrumental response functions P7SOURCE\_V6 were used. The lower energy bound was set at 100\,MeV and the region of interest (ROI) radius at 10$^\circ$ \citep[see][]{abd09}. All the nearby sources in the 2FGL catalog \citep{nol12} were included in the model of the ROI,
fixing their spectral index and letting their normalization free in the fit. Since no variability is expected for the underlying background diffuse emission, we fixed the
normalization of both Galactic and extragalactic diffuse backgrounds to the average values determined from
the whole observing period described in this paper (from August 4, 2008 to February 21, 2011).
The early portion of the data between August 4, 2008 and August 13, 2009 coincides with the early LAT observations of NGC\,1275 presented in \citet{kat10} \citep[see also][]{bro11}.
The systematic uncertainties on the flux were estimated as 10\% at 100\,MeV, decreasing to 5\% at 560\,MeV, and increasing to 10\% at 10\,GeV and higher \citep{ack12}.

\begin{figure*}
\centering
\includegraphics[width=0.85\textwidth]{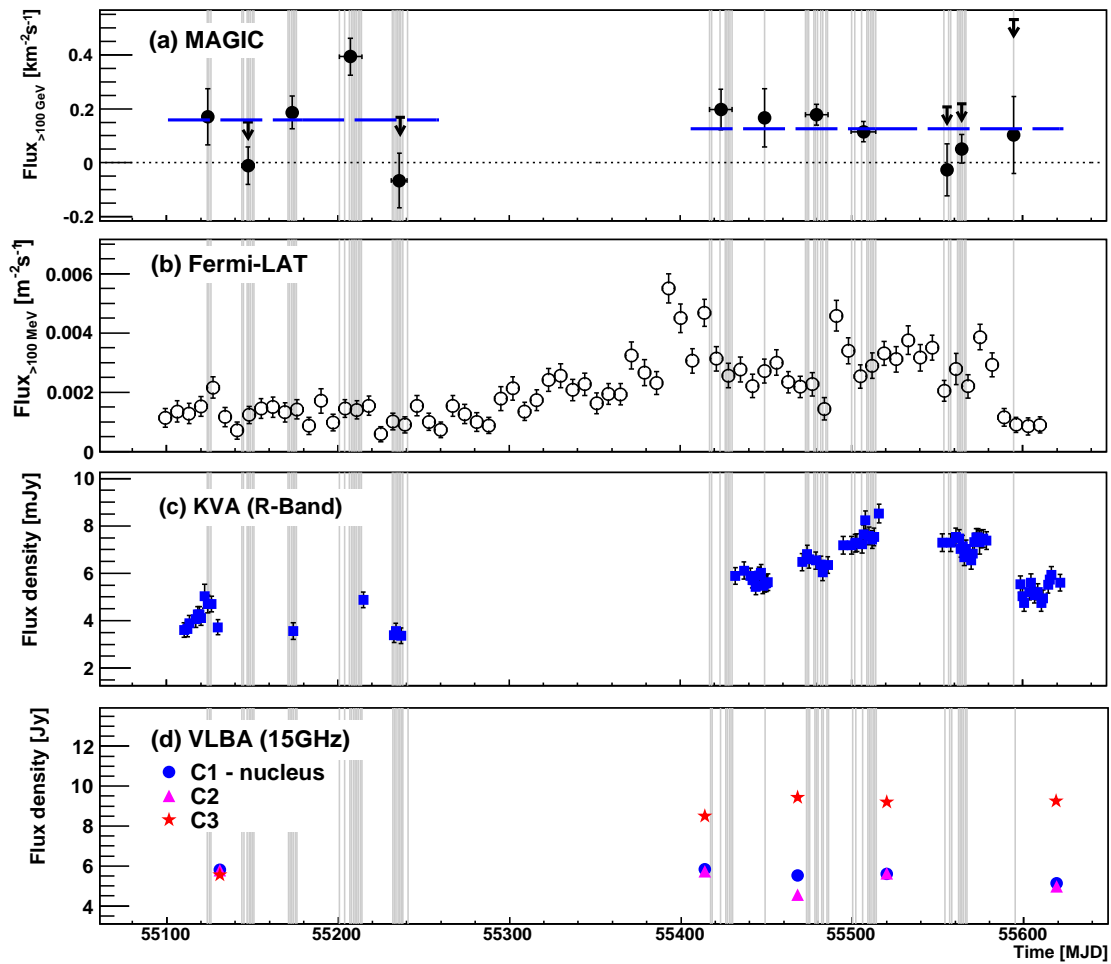}
\caption{
Multi-frequency light curves of NGC\,1275 from October 2009 to February 2011.
The vertical gray lines show the exact observation dates with MAGIC.
(a)~The MAGIC LC above $100$\,GeV in monthly bins.
The thick dashed lines represent the constant fit for both observation campaigns.
For data showing an excess $<$1\,$\sigma$, upper limits (black arrows) are calculated assuming a spectral index
$\Gamma$=4.0, using the \citealp{rol05} method with a confidence level of $95\%$ and a total systematic uncertainty of 30$\%$.
(b)~The \textit{Fermi}-LAT LC above 100\,MeV has a weekly binning.
(c)~The KVA R-band LC shows the core continuum flux corrected for Galactic extinction. The constant systematic error ($\pm$0.6\,mJy) induced by the host-galaxy and emission-line subtraction is not included in the error bars.
(d)~The VLBA LCs show the 15\,GHz emission from the three innermost components of the radio jet.
}
\label{fig:MWL_LC}
\end{figure*}

Panel~(b) of Fig.~\ref{fig:MWL_LC} shows the $\gamma$-ray flux above 100\,MeV of NGC\,1275 from September 25, 2009 to February 18, 2011 binned at one-week intervals. In each bin, the significance of the NGC\,1275 detection is higher than the test statistic TS$>$10 and the ratio of flux error to flux is below 0.5 \citep[$\delta$F/F\,\lapp\,0.5, see][]{nol12}. There is a general trend of increasing flux toward the end of the observations, and several episodes of large flares with a flux variability as fast as one week.

\begin{figure}
\centering
\includegraphics[width=0.47\textwidth]{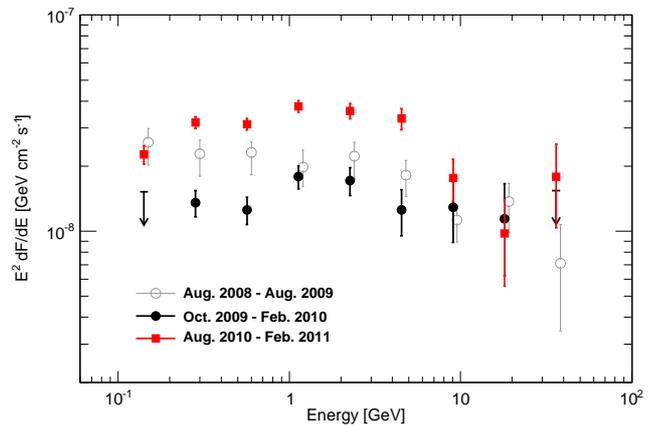}
\caption{NGC\,1275 SED measured with \textit{Fermi}-LAT during several periods corresponding to the first year of operation \citep[empty circles,][]{kat10}, the first MAGIC campaign (full circles and black arrows), and the second MAGIC campaign (full squares).}
\label{fig:Fermi_SED}
\end{figure}

Fig.~\ref{fig:Fermi_SED} shows the \textit{Fermi}-LAT spectral energy distributions (SED) derived for the two MAGIC-observation campaigns
(MJD 55123--55241 for Camp.\,1 and MJD 55417--55595 for Camp.\,2) compared with that obtained for the first year
given in \cite{kat10}. The spectra were obtained from nine independent energy bins: (0.1--0.2, 0.2--0.4, 0.4--0.8, 0.8--1.6, 1.6-3.2, 3.2--6.4, 6.4--12.8, 12.8--25.6, and 25.6--51.2)\,GeV.
The general spectral shape can be approximated by a power-law function with a photon index $\Gamma \simeq 2$, slightly curving down toward the highest energies.
Substantial spectral evolution can be seen in different observational periods, both in the flux and peak frequencies of the $\gamma$-ray emissions.
The Camp.\,2 mean GeV flux is about twice as high as during the first MAGIC campaign, and
the Camp.\,2 SED shows a clear curvature peaking around 1\,GeV, whereas the first-year SED is constantly decreasing with energy.
However, above a few GeV all the spectra reach similar fluxes. The $\gamma$-ray flux seems more variable at low energy. 
The precise fit characterization of the LAT spectrum is discussed together with the MAGIC data in Section~\ref{subsec:3.3}.

\subsection{\textit{Chandra X-ray Observatory}}
\label{subsec:2.3}
The ACIS detector onboard the \textit{Chandra X-ray Observatory} is a spectrometer with very high angular resolution ($\sim$0.5$\arcsec$) operating in the range of 0.2--10\,keV. In the public archive, we found seven observations performed during the MAGIC observing period
(namely ObsId 11713, 11714, 11715, 12025, 12033, 12036, and 12037 ; PI Fabian).
In NGC\,1275, the nuclear X-ray flux is high enough that pileup is potentially a severe problem.
For an on-axis observation, PIMMS estimates a total pileup fraction of about 52\%
(i.e. more than half of the incident photons are either combined into piled events or removed because their energy is  higher than the spacecraft threshold).
The nucleus was not the primary target of these observations and no observing strategy for minimizing nuclear pileup was actuated
(e.g. selecting a subarray to reduce the nominal frame integration time).
However, this set of observations was centered several arc-minutes away from the nucleus
(3.2$\arcmin$--7.6$\arcmin$), and this reduces the pileup because at larger offset the effective
area of the mirrors is lower and the point-spread function (PSF) is larger.
We analyzed the three observations with offset angles $>$7.5\,$\arcmin$ (ObsId 12025, 12033, 12036) that are less affected by the pileup. The measured counts to frame are still higher than the typical threshold \citep[0.1-0.2, e.g.][]{mas12} above which the pileup is a problem. Thus, a pileup model \citep{dav01} is statistically required to fit the data.

\begin{figure*}
\centering
\includegraphics[width=0.49\textwidth]{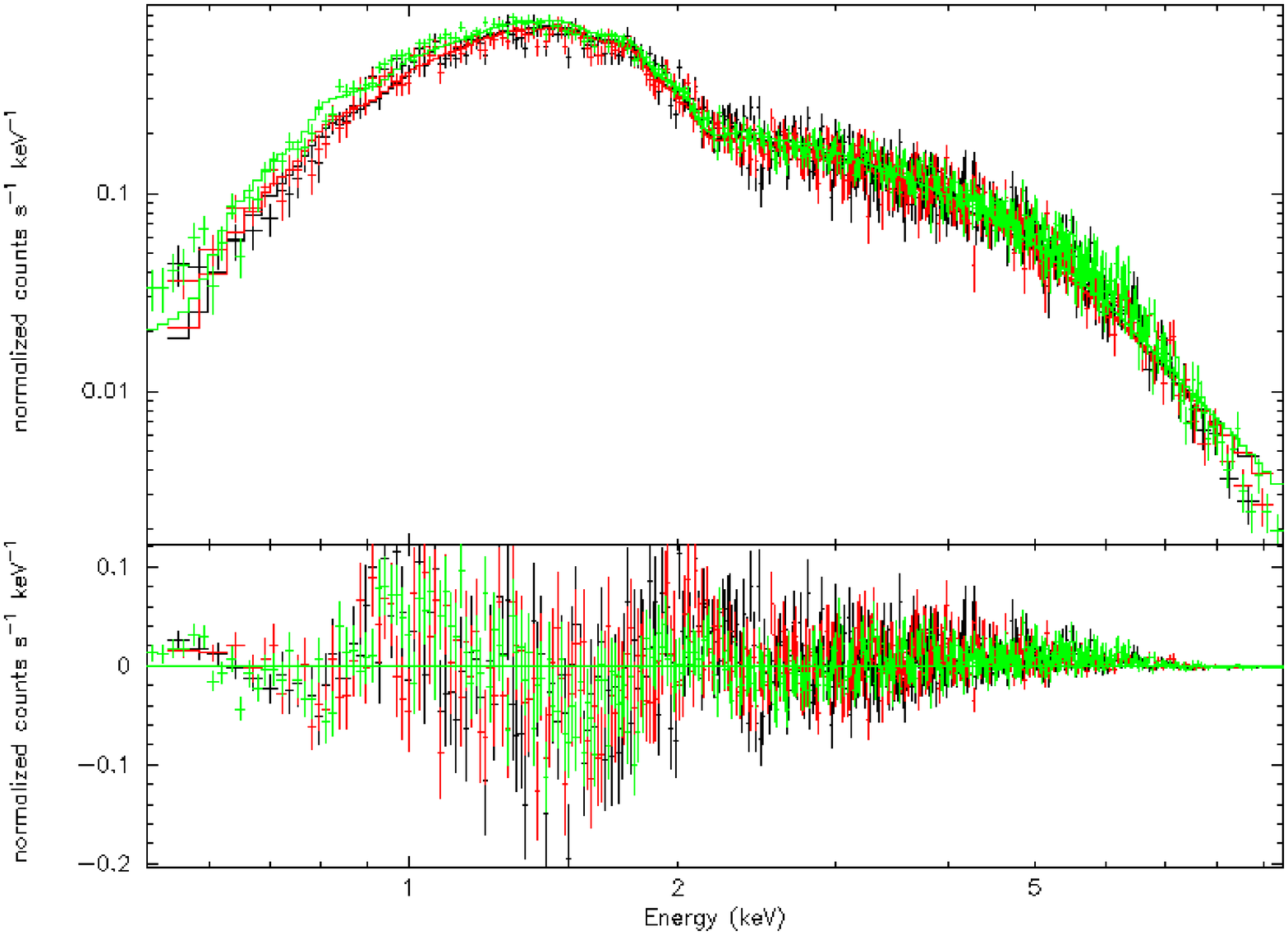}
\includegraphics[width=0.46\textwidth]{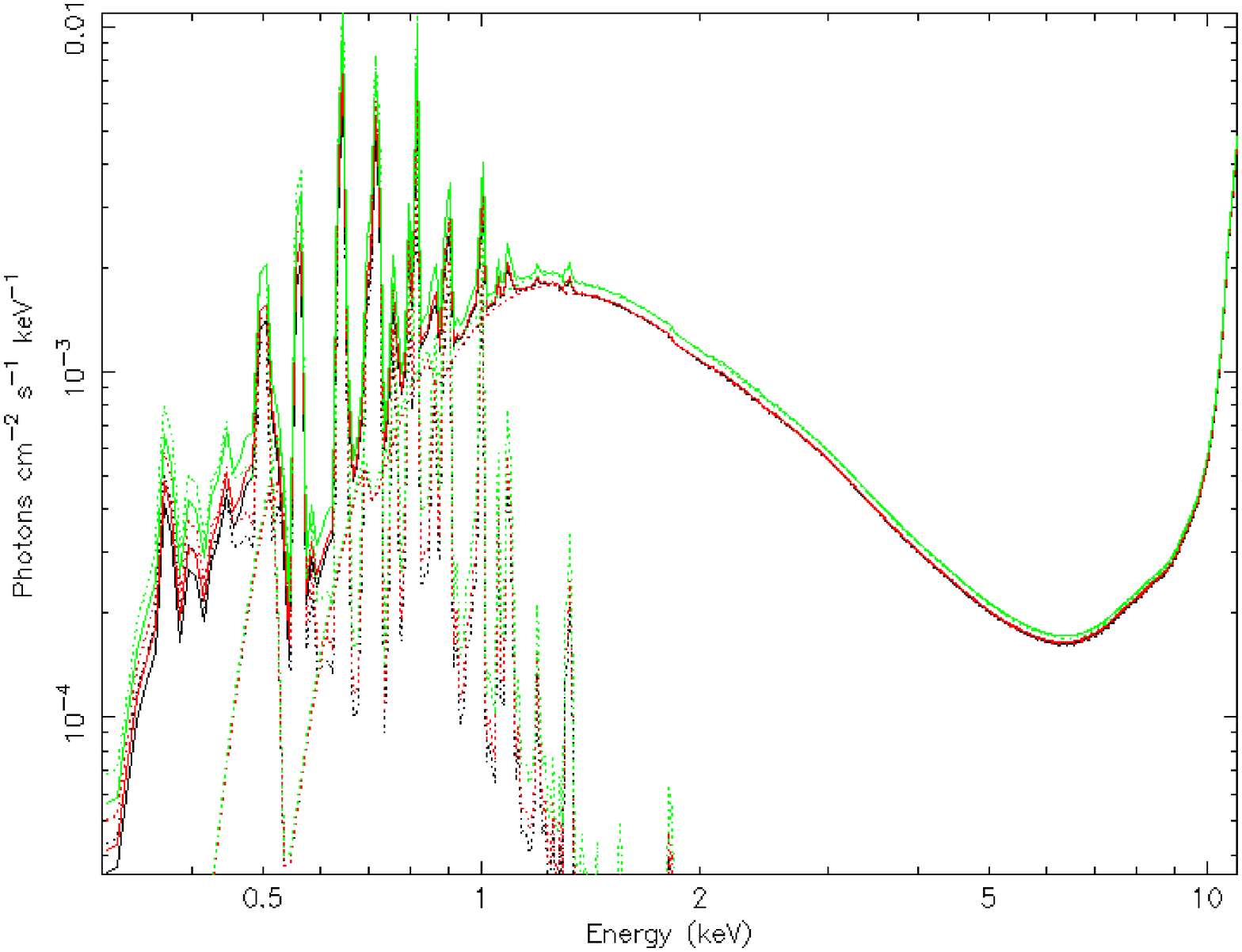}
\caption{
Energy spectra of the NGC\,1275 nuclear emission from the three large offset \textit{Chandra} observations (ObsId 11025-green, 12033-red, 12036-black) and the fit models with fixed power-law index $\Gamma=2.5$. In the top left panel, crosses show the photon rates per energy bin grouped to have at least 25 counts/bin and the solid lines show the fit models. In the bottom left panel, the fit residuals are reported. The right-hand panel shows the photon flux energy spectra of the fit models. Dashed lines represent the different model components (\textit{mekal} and \textit{powerlaw} after absorption and pileup effect) and the solid lines are the total. The large tail at high energy is a consequence of the pileup modeling.
}
\label{fig:Chandra_spectra}
\end{figure*}

We applied the standard data reduction procedure\footnote{http://cxc.harvard.edu/ciao/guides/acis\_data.html} using the \textit{Chandra} Interactive Analysis of Observations CIAO~4.4, with \textit{Chandra} Calibration Database CAL-DB version 4.4.1.
We generated a new level = 2 event file by applying the standard grade, status, and good time filters.
We rebinned the spectrum using a 25-count threshold and fitted the spectrum in the range 0.5--9.5\,keV.
The three spectra were fitted simultaneously by adopting the model $pileup\times phabs(mekal+zphabs(powerlaw))$ in XSPEC version: 12.7.1,
where $mekal$ models the thermal and line emissions of the hot diffuse gas, $powerlaw$ is the non-thermal power-law emission from the jet, $pileup$ reproduces the pileup effect and, $phabs$ and $zphabs$ correspond to the Galactic and internal absorptions.
The metal abundance and hydrogen density of $mekal$ were fixed to 0.7 solar value and 0.1\,cm$^{-3}$, respectively.
The temperature and normalization are obtained from the fit ($kT=0.30\pm0.01$\,keV).
For the absorption, we fixed the hydrogen column density from the Galaxy ($N_{H,gal}$=1.5$\times$10$^{21}$\,cm$^{-2}$) and let free the internal absorption ($zN_H$=5.6$\pm$0.2$\times$10$^{21}$\,cm$^{-2}$).

The power-law slope $\Gamma$ is strongly affected by the pileup effect, which reduces the counts at lower energy (because two or more photons are counted as one) and increases the counts at high energy. This creates a kind of degeneracy, and we decided to fix the parameter $\Gamma$ in the model.
All the parameters of the $powerlaw$ and $mekal$ models were linked, except for the normalization.
Table~\ref{tab:Xrayflux} provides the results of the fit for several power-law slope assumptions \textbf{and
Fig.~\ref{fig:Chandra_spectra} shows the measured energy spectra and the fit models for $\Gamma=2.5$}.

\begin{table}[htdp]
\caption{Unabsorbed integral flux of the NGC\,1275 non-thermal power-law core emission in the range 2-10\,keV obtained from \textit{Chandra} observations assuming different power-law slopes.}
\begin{center}
\begin{tabular}{lccccc}
\hline
\hline
ObsId & Date & $\Gamma$ & fit $\chi ^2$ (d.o.f.) & F(2-10\,keV)\tablefootmark{a}\\
\hline
12025 & 2009-11-25 &     &               & 1.26 \begin{footnotesize}$^{+0.1} _{-0.1}$ \end{footnotesize}  \\
12033 & 2009-11-27 & 2.5 & 1616.8 (1075) & 1.27 \begin{footnotesize}$^{+0.1} _{-0.1}$ \end{footnotesize} \\
12036 & 2009-12-02 &     &               & 1.35 \begin{footnotesize}$^{+0.1} _{-0.1}$ \end{footnotesize}  \\
\hline
12025 & 2009-11-25 &     &               & 1.50 \begin{footnotesize}$^{+0.1} _{-0.1}$ \end{footnotesize}  \\
12033 & 2009-11-27 & 2.0 & 1307.4 (1075) & 1.49 \begin{footnotesize}$^{+0.5} _{-0.1}$ \end{footnotesize}   \\
12036 & 2009-12-02 &     &               & 1.71 \begin{footnotesize}$^{+0.4} _{-0.2}$ \end{footnotesize}  \\
\hline
12025 & 2009-11-25 &     &               & 2.20 \begin{footnotesize}$^{+0.3} _{-0.3}$ \end{footnotesize} \\
12033 & 2009-11-27 & 1.7 & 1392.8 (1075) & 4.40 \begin{footnotesize}$^{+1.2} _{-0.3}$ \end{footnotesize} \\
12036 & 2009-12-02 &     &               & 5.60 \begin{footnotesize}$^{+0.1} _{-0.2}$ \end{footnotesize} \\
\hline
\end{tabular}
\tablefoottext{a}{Integral flux in 10$^{-11}$\,erg\,cm$^{-2}$\,s$^{-1}$ (uncertainties at $\pm$1$\sigma$)} \\
\end{center}
\label{tab:Xrayflux}
\end{table}%

\subsection{KVA and NOT}
\label{subsec:2.4}

NGC\,1275 has been monitored in the optical R-band by the Tuorla blazar monitoring program\footnote{http://users.utu.fi/kani/1m} since October 2009. The observations are performed using the KVA 35\,cm telescope located at La Palma. The data were reduced using the standard data analysis pipeline and the magnitudes were measured with differential photometry with an aperture of 5.0$\arcsec$ and the comparison stars from \cite{fio98}. The observed magnitudes varied between 13.05 and 13.3. The magnitudes were converted to flux using the standard formula $S=S_0\cdot10^{-mag/2.5}$ and $S_0=3080$\,Jy.

To calculate the intrinsic emission from the core, the measured magnitudes over the MAGIC observing periods have to be averaged, the host galaxy subtracted, and the resulting data de-reddened. Additionally, the core flux is contaminated by emission lines that also need to be subtracted. To estimate the contributions of the host galaxy and the emission lines to the raw KVA measurements, NGC\,1275 was observed with the Nordic Optical Telescope (NOT) on November 15, 2011 using the ALFOSC instrument. Five 60\,s R-band exposures and two spectra using grism 4 and slit 1$\arcsec$ were obtained. The images were reduced and combined using the standard IRAF tasks. The methodology of \cite{nil07} was used to estimate the contribution of the host galaxy. In short, the method is based on fitting a two-dimensional nucleus plus host galaxy model to the image and convolving the model with the PSF derived from the stars in the same field of view. The resulting host galaxy flux based on the aperture used is 
$11.08\pm0.55$\,mJy. The line contamination is estimated from the spectrum of NGC\,1275 by integrating the observed line flux within the NOT R-band filter. We find that the line contamination in the R-band is 0.9\,mJy, that is only $\sim$10\% of the host flux. The host and line contaminations were then subtracted to derive the core continuum non-thermal flux, which was corrected for Galactic extinction (de-reddened) using the extinction value from \cite{sch98}.
The measured fluxes are shown together with the multi-frequency light curves (Fig.~\ref{fig:MWL_LC}).
The mean fluxes over the two MAGIC campaigns are (4.1$\pm$0.6)\,mJy for Camp.\,1 and (6.5$\pm$0.6)\,mJy for Camp.\,2.




\subsection{MOJAVE observations}
\label{subsec:2.6}

\begin{figure}
\begin{center}
\includegraphics[width=0.245\textwidth]{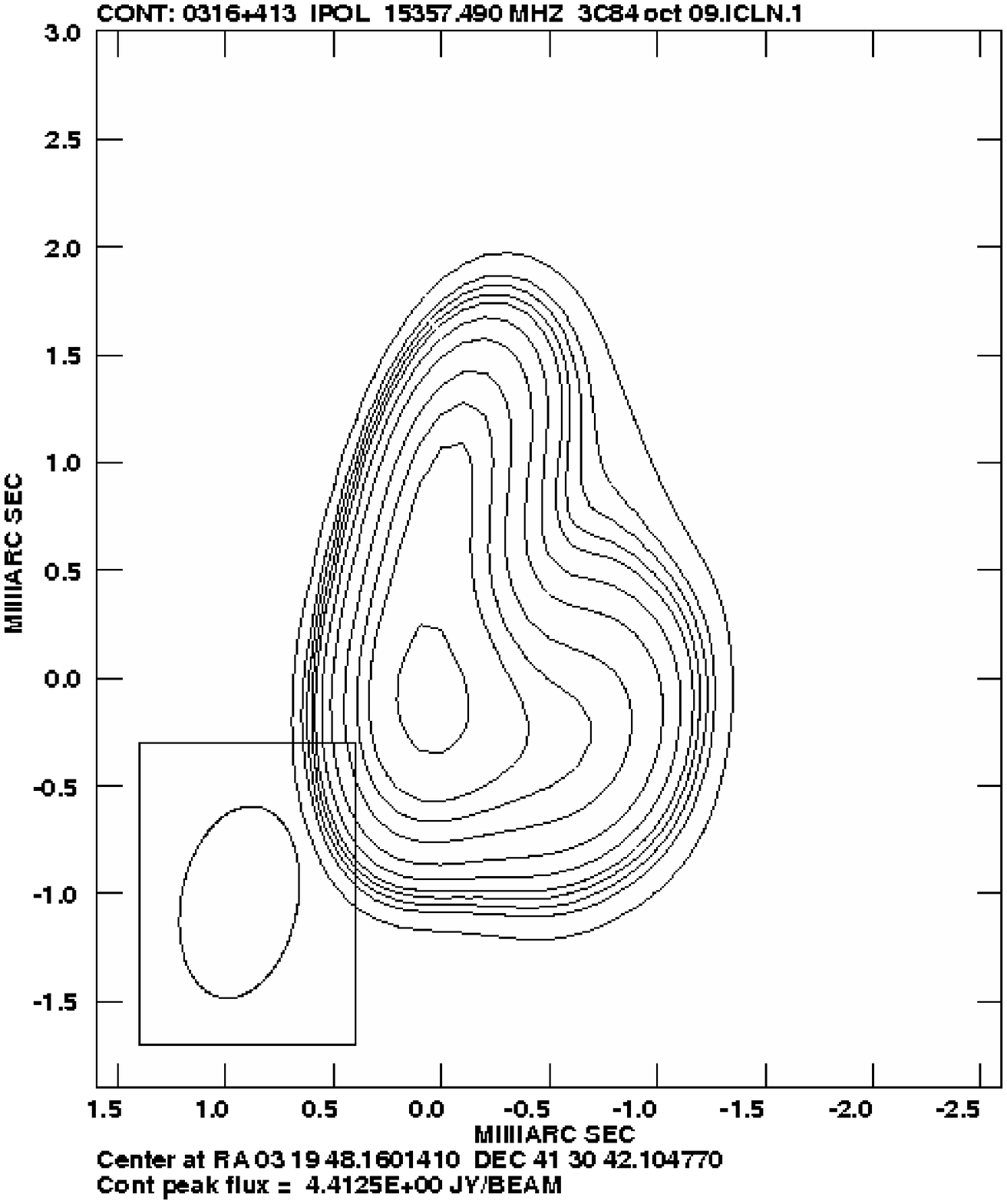}\includegraphics[width=0.245\textwidth]{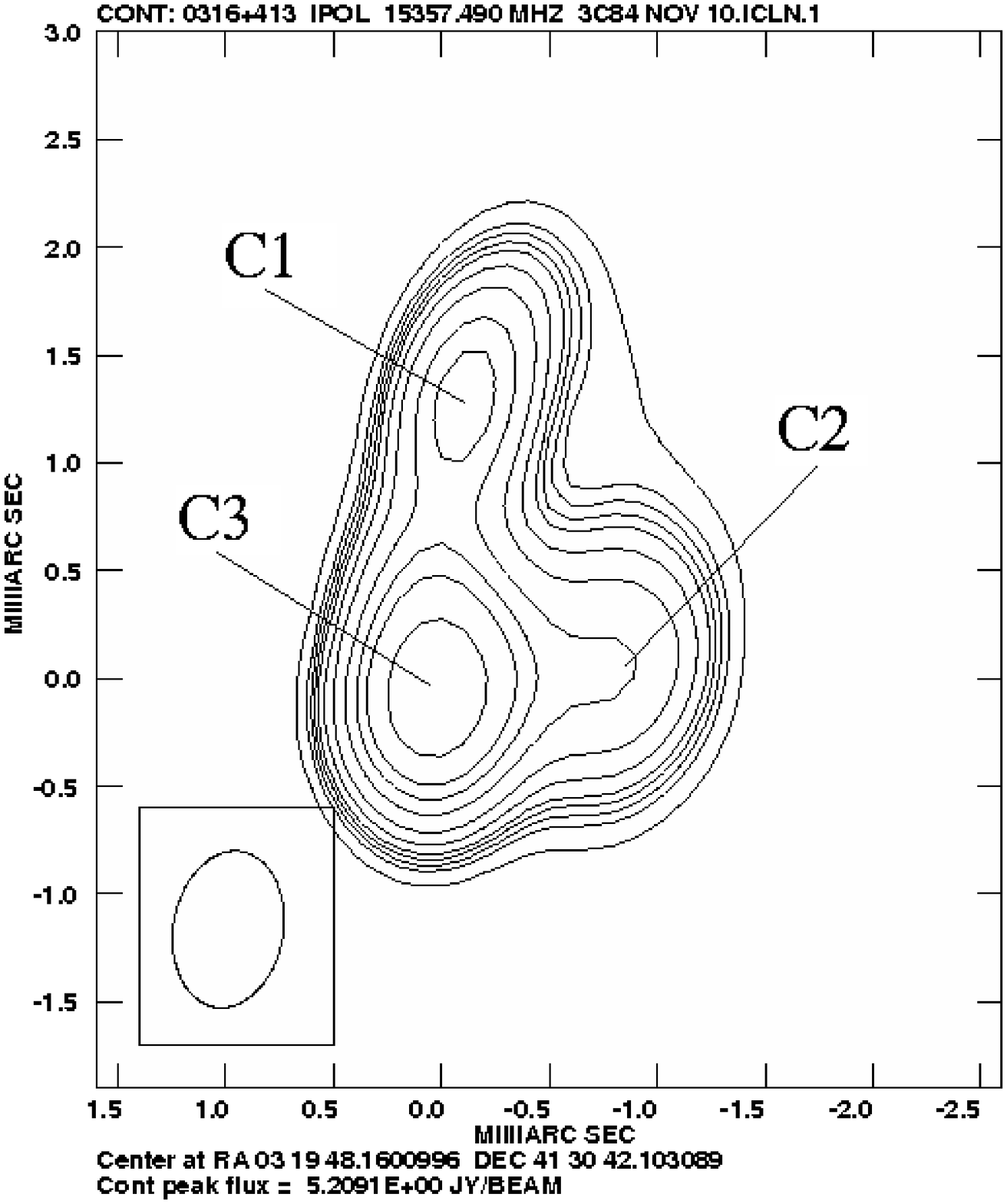}
\caption{MOJAVE images of 3C\,84 at 15\,GHz in October 2009 (left) and November 2010 (right).
The contours are plotted at levels of (0.5, 0.7, 0.8, 0.9, 1, 1.2, 1.5, 2, 2.5, 3, and 4)\,Jy/beam.
The labels C1, C2, and C3 show the three bright components in the central region. The insets show the shape of the PSF.
}
\label{fig:maps}
\end{center}
\end{figure}

The Very Long Baseline Array (VLBA) is a radio interferometer using ten antennas 25\,m in diameter located across the USA between Hawaii and the Virgin Islands. Thanks to its very long baselines (up to 8600\,km), angular resolutions down to milli-arcseconds (mas) can be achieved. The Monitoring Of Jets in AGNs with VLBA Experiments (MOJAVE) program is constantly observing a set of bright AGNs in the northern hemisphere at 15\,GHz \citep{lis09}.

The calibrated data of the MOJAVE program were reduced using the NRAO Astronomical Imaging Processing System (AIPS). In this work, we analyzed five epochs covering the time interval from October 2009 to February 2011.
Fig.~\ref{fig:maps} shows the total intensity images of the central region ($\sim$1\,pc) at 15\,GHz, featuring three bright components C1, C2, and C3. Generally C1 is understood to be the nuclear core emission. C3 corresponds to the rapidly growing component that recently appeared \citep{nag10}. The position and flux density for each component was derived by means of the task JMFIT, which fits a Gaussian function to a selected area in the image plane. The obtained flux densities are listed in Table~\ref{tab:radioflux} and are shown in panel (d) of Fig.~\ref{fig:MWL_LC}.
We consider the VLBA accuracy of amplitude calibration as the dominant uncertainty (at the level of 5\%).
While the flux densities of C1 and C2 do not show considerable variability, the intensity of C3 significantly increases between the beginning and the end of the observational campaign (as reported Table~\ref{tab:radioflux}).

\begin{table}[htdp]
\caption{Flux densities of components C1, C2, and C3 obtained from MOJAVE data at 15\,GHz.}
\begin{center}
\begin{tabular}{lccc}
\hline
\hline
Epoch & C1 (Jy) & C2 (Jy) & C3 (Jy) \\
\hline
2009-10-09 & 5.80 $\pm$ 0.29 & 5.79 $\pm$ 0.29 & 5.54 $\pm$ 0.27 \\
2010-08-06 & 5.84 $\pm$ 0.29 & 5.74 $\pm$ 0.29 & 8.50 $\pm$ 0.43 \\
2010-09-29 & 5.48 $\pm$ 0.27 & 4.53 $\pm$ 0.23 & 9.42 $\pm$ 0.47 \\
2010-11-20 & 5.59 $\pm$ 0.28 & 5.62 $\pm$ 0.28 & 9.20 $\pm$ 0.46 \\
2011-02-27 & 5.13 $\pm$ 0.26 & 4.97 $\pm$ 0.25 & 9.21 $\pm$ 0.46 \\
\hline 
\end{tabular}
\end{center}
\label{tab:radioflux}
\end{table}%

\subsection{Other multi-frequency data}
\label{subsec:2.7}
NGC\,1275 is a famous and bright object that is regularly observed at many frequencies.
Some observations made during the period of interest here (October 2009 -- February 2011) were reported in scientific journals
and can be used for our broadband study without dedicated data processing.

NGC\,1275 was part of the Early Release Compact Source Catalogue of the \textit{Planck} microwave observatory \citep[and references therein]{pla11}.
This catalog contains NGC\,1275 observations taken during two periods: from August 29 to September 4, 2009
and from February 10 to 19, 2010. Both periods are close to the first MAGIC observation campaign.
More recent \textit{Planck} observations taken from August 29 to September 4, 2010, corresponding to the second MAGIC campaign,
are reported in \cite{gio12}. The derived spectrum between 30\,GHz and 857\,GHz agrees well with the one reported in the Early Release Compact Source Catalogue between 30\,GHz and 353\,GHz. The microwave flux of NGC\,1275 was quite stable during the MAGIC observations.

\cite{gio12} also reported contemporaneous observations in the ultraviolet\,(UV)/optical band (170--650\,nm) with \textit{Swift}-UVOT taken August 9, 2010 at the beginning of the second MAGIC campaign.
\textit{Swift}-XRT cannot easily be used to estimate the nuclear X-ray emission because its angular resolution does not allow isolation of the AGN from diffuse Perseus cluster central emission. The analysis of \textit{Swift}-XRT and -BAT data reported in \cite{aje09} shows that the cluster emission dominates until $\sim$30\,keV. NGC\,1275 X-ray emission is also monitored with \textit{Swift}-BAT\footnote{http://swift.gsfc.nasa.gov/results/bs70mon/SWIFT J0319.7p4132}. The light curve (dominated by the cluster emission) does not show any enhancement during the MAGIC observation period, suggesting that no exceptional hard X-ray flare occurred.

In radio, NGC\,1275 was regularly monitored with the Mets\"ahovi 14\,m single-dish telescope with additional VLBI observations of the VERA stations at 43\,GHz \citep{nag12} and the MOJAVE program at 15\,GHz (discussed in Section~\ref{subsec:2.6}).
Moreover, the GENJI program \citep{nag12b} started a dense VLBI monitoring of NGC\,1275 at 22\,GHz in November 2010 (during the second MAGIC observation campaign). All radio data show the same trend of a flux slowly increasing with time from the component C3 during the period of interest.


\section{Interpretations}
\label{sec:3}

\subsection{Data contamination}
\label{subsec:3.1}
The majority of known VHE $\gamma$-ray emitting AGNs are blazars, that is, their jet points directly to the observer.
In these cases, the non-thermal jet emission is highly boosted and dominates other components.
It is rather simple to subtract the background emission from the host galaxy.
For NGC\,1275, however, this is much more difficult. Thermal and non-thermal radiations can be emitted by extended structures in the host galaxy or even the host cluster of galaxies. Thus, observations with different angular resolution can integrate signal from different regions.

While the rapid and strong variability measured by \textit{Fermi}-LAT implies that the $\gamma$-ray emission is largely dominated by a compact source most likely close to the nucleus, there are strong indications that at other frequencies other emitting components contribute significantly to the observed flux.
In optical, the KVA measurements were corrected for both the host-galaxy and emission-line contributions.
They are still significantly higher than previous \textit{Hubble Space Telescope} (\textit{HST}) measurements taken in 1994--1995 \citep{chiab99,baldi}
when the radio core emission was similar to its level in 2009--2011 \citep[see Effelsberg light curve in][]{nag10}.
The HST observations are characterized by an extremely good angular resolution, allowing the emission to be pinpointed from the innermost regions of the jet. The small discrepancy between our measurement and the \textit{HST} results might be explained by temporal variability, but also by the limited angular resolution of KVA, which may contain large-scale jet contributions.
In radio, VLBA provides an extremely good angular resolution which allows us to resolve three components in the sub-pc core region. However, this resolution ($\sim$0.2\,pc) is still insufficient to probe the $\gamma$-ray emission-region size expected from the week-scale variability ($\sim$0.01\,pc). Sub-structures not resolved by VLBA can therefore exist.
\textit{Planck} and UVOT both have relatively broad angular resolution and their measurements are most likely contaminated by large-scale emission.
Thus our results from radio to X-rays should be considered firstly as upper limits of the low-energy counterpart of the $\gamma$-ray emitting region.

\subsection{Optical $\gamma$-ray correlation}
\label{subsec:3.2}
\begin{figure}
\centering
\includegraphics[width=0.44\textwidth]{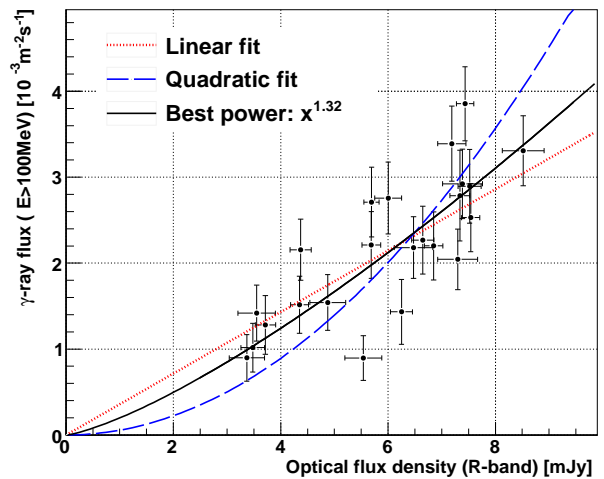}
\caption{NGC\,1275 $\gamma$-ray flux above 100\,MeV (\textit{Fermi}-LAT) as a function of the 
optical continuum core flux density (KVA). The dotted and dashed lines represent the linear and quadratic fit, respectively. The solid line is the best fit with a free correlation factor ($F_{\gamma} = \alpha (F_{opt})^\beta$ with $\beta$=1.32).}
\label{fig:GeV-Opt_Cor}
\end{figure}

We measured flux variability at different energy bands (Fig.~\ref{fig:MWL_LC}). The light curves for GeV $\gamma$-rays, optical, and the C3 radio component all show a similar trend. The intense monitoring in optical and GeV $\gamma$-rays allows us to study this correlation more deeply.
For every time bin of the \textit{Fermi}-LAT weekly LC containing simultaneous KVA observations, we calculated the mean optical flux density of the core. Fig.~\ref{fig:GeV-Opt_Cor} shows the $\gamma$-ray flux above 100\,MeV as a function of the optical core emission for the 23 independent weeks with simultaneous \textit{Fermi}-LAT and KVA observations.
A clear correlation is visible. The linear correlation coefficient (Bravais-Pearson coefficient) is 0.79$^{+0.07}_{-0.10}$.
The correlation can also be characterized using the prescription of \citet{ede88}, which, unlike the Pearson coefficient calculation, takes into account the measurement uncertainties. This leads to a discrete correlation function DCF\,=\,0.86$\pm$0.21. Both statistical procedures for quantifying the optical $\gamma$-ray correlation lead to the same conclusion: a positive correlation at the level of 4--5 $\sigma$.

This correlation strongly suggests that the corrected KVA results are dominated by the optical counterpart
of the $\gamma$-ray-emitting region.
The lower optical level measured in 1994--1995 by \textit{HST} ($<$2\,mJy) is most likely due to a weaker inner jet activity
which, indeed, was not detected in $\gamma$-rays with \textit{CGRO}-EGRET during this period.

Different types of correlation can be expected depending on the emission model and on the physical parameter driving the flux variation. In a simple SSC model (see description Section~\ref{subsec:3.4}), the variation of the electron density implies a quadratic correlation, whereas the magnetic field variation implies a linear correlation.
We fitted the data assuming correlation functions of the form $F_{\gamma} = \alpha (F_{opt})^\beta$.
The best fit is obtained for $\beta$=1.32, but both a linear ($\beta$=1) and a quadratic ($\beta$=2) correlation can fit the data as well.

\subsection{Fit of $\gamma$-ray SED}
\label{subsec:3.3}
\begin{figure}
\centering
\includegraphics[width=0.5\textwidth]{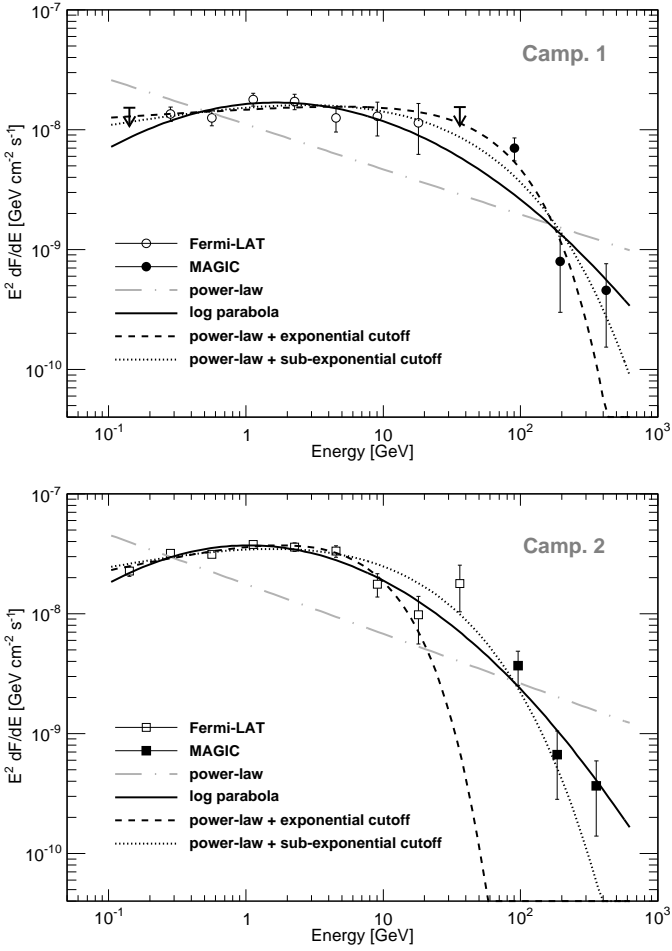}
\caption{NGC\,1275 SED in $\gamma$-ray band measured with MAGIC and \textit{Fermi}-LAT together with fit functions of the data (see parameters in Table~\ref{t_fitPar}). Upper and lower panels correspond to the periods October 2009 -- February 2010 and August 2010 -- February 2011, respectively.}
\label{fig:Gamma_SED}
\end{figure}

The photon indices of the energy spectrum measured quasi-simultaneously with \textit{Fermi}-LAT ($\Gamma\simeq$2) and MAGIC ($\Gamma\simeq$4) are very different. Fig.~\ref{fig:Gamma_SED} shows the SED between 100\,MeV and 650\,GeV corresponding to the two observation campaigns.
We fitted several functions to these SEDs, taking into account only the statistical errors of each data point.
Table~\ref{t_fitPar} gives the functional forms used and the fit results.

\begin{table}[h]
\caption{\label{t_fitPar} Parameters of the NGC\,1275 $\gamma$-ray SED fit over the energy range 0.1 -- 650\,GeV, for different fit functions:}
\centering

\begin{tabular}{l}
\begin{small} power law: $dF/dE = f_0 (\frac{E}{GeV})^{-\Gamma}$  \end{small}\\
\begin{tabular}{lcccc}
\hline\hline
Epoch & $f_0$\tablefootmark{a} & $\Gamma$ & $\chi^2$/dof & Prob.\\
\hline
Camp.\,1  &111$\pm$8 & 2.38$\pm$0.03 & 57.4/8 & 1.5e-9\\
Camp.\,2  &176$\pm$7 & 2.41$\pm$0.02 & 317/10 & 3.e-62\\
\hline
\end{tabular} \\
\\
power law with exponential cutoff:
$\frac{dF}{dE} = f_0 (\frac{E}{GeV})^{-\Gamma} e^{-E/E_c}$\\
\begin{tabular}{lccccc}
\hline\hline
Epoch & $f_0$\tablefootmark{a} & $\Gamma$ & $E_c$\tablefootmark{b}  & $\chi^2$/dof & Prob.\\
\hline
Camp.\,1  &149$\pm$10 & 1.93$\pm$0.06 & 67$\pm$12& 8.6/7 & 0.28\\
Camp.\,2  &411$\pm$25 & 1.75$\pm$0.05 & 7.3$\pm$1.6& 26.5/9 & 0.0017\\
\hline
\end{tabular}\\
\\
power law with a sub-exp. cutoff:
$\frac{dF}{dE} = f_0 (\frac{E}{GeV})^{-\Gamma} e^{-\sqrt{E/E_c}}$\\
\begin{tabular}{lccccc}
\hline\hline
Epoch & $f_0$\tablefootmark{a} & $\Gamma$ & $E_c$\tablefootmark{b}  & $\chi^2$/dof & Prob.\\
\hline
Camp.\,1  &201$\pm$17 & 1.77$\pm$0.08 & 13.1$\pm$4.2& 9.4/7 & 0.22\\
Camp.\,2  &536$\pm$51 & 1.72$\pm$0.05 & 5.0$\pm$1.6& 19.3/9 & 0.023\\
\hline
\end{tabular}\\
\\
log parabola:
$dF/dE = f_0 (\frac{E}{GeV})^{- 2-\beta log(E/E_p)}$ \\
\begin{tabular}{lccccc}
\hline\hline
Epoch & $f_0$\tablefootmark{a} & $\beta$ & $E_p$\tablefootmark{c}  & $\chi^2$/dof & Prob.\\
\hline
Camp.\,1  &163$\pm$11 & 0.26$\pm$0.04 & 2.8$\pm$1.5& 12.3/7 & 0.09\\
Camp.\,2  &371$\pm$13 & 0.31$\pm$0.03 & 1.1$\pm$0.2& 13.4/9 & 0.15\\
\hline
\end{tabular}

\end{tabular}
\tablefoottext{a}{Differential flux normalization in 10$^{-10}$\,GeV$^{-1}$\,cm$^{-2}$\,s$^{-1}$}\\
\tablefoottext{b}{Cutoff energy in GeV}\\
\tablefoottext{c}{Peak energy in GeV}
\end{table}


A power-law fit is completely excluded, as suggested by previous VHE upper limits contemporaneous to \textit{Fermi}-LAT data in 2008 \citep{acc09,ale10b}. A power-law spectrum with an exponential cutoff fits the first campaign well but does not match the second one, where the LAT data clearly show a curvature. Assuming a softer cutoff (a power-law with a sub-exponential cutoff), one can fit both campaigns quite well. This gives a photon index of $\Gamma\simeq$\,1.75 at low energy, which matches well the non-thermal power-law component seen in X-ray by \textit{Chandra} \citep{bal06} and \textit{Swift}-BAT \citep{aje09}. As explained in \citet{acc09}, this function corresponds to the $\gamma$-ray spectral shape expected for the inverse Compton emission of an electron population with an energy distribution following a power law with an exponential cutoff at high energy\footnote{Strictly speaking, a sub-exponential description for the cutoff of the inverse Compton spectrum is valid only in the Thomson regime. 
As discussed in Section~\ref{subsec:3.4}, the scattering responsible for emission at the high-energy peak in NGC\,1275 mainly occurs in the Klein-Nishina regime. In this case, that particular shape can still be used as an approximation.}.

Finally, a curved power-law (log parabola) provides the best fit for Camp.\,2 data and a reasonably good fit for Camp.\,1. It allows us to determine the peak energy of the SED, which varies marginally from 2.8$\pm$1.5\,GeV (Camp.\,1) to 1.1$\pm$0.2\,GeV (Camp.\,2).

\subsection{Broadband SED and emission models}
\label{subsec:3.4}

\begin{figure}
\centering
\includegraphics[width=0.45\textwidth]{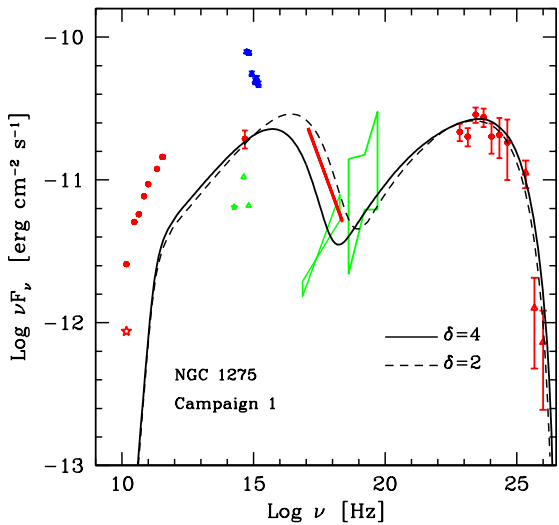}
\caption{NGC\,1275 SED for the epoch of Camp.\,1 (October 2009 -- February 2010). Red symbols show nearly simultaneous data from MOJAVE, KVA, \textit{Chandra} (when assuming a power-law slope $\Gamma$=2.5), \textit{Fermi}-LAT, and MAGIC. Contemporaneous \textit{Planck} data from the \cite{pla11} are also shown with red circles. The red star in radio shows the level of the individual sub-pc components (C1, C2, and C3 have nearly the same flux). Green bow-ties in the X-ray band report the archival \textit{Chandra} and \textit{Swift}-BAT results from \cite{bal06} and \cite{aje09}, respectively. Green symbols report optical-IR measurement by \textit{HST} from \cite{chiab99} (triangles) and \cite{baldi} (star). The blue points show the August 2010 \textit{Swift}-UVOT data from \cite{gio12} (corrected for the host galaxy contribution). The dashed and solid lines report the SSC models for $\delta=2$ ($\theta_v<30^\circ$) and $\delta=4$ ($\theta_v<15^\circ$), respectively.}
\label{fig:SED5}
\end{figure}

\begin{figure}
\centering
\includegraphics[width=0.45\textwidth]{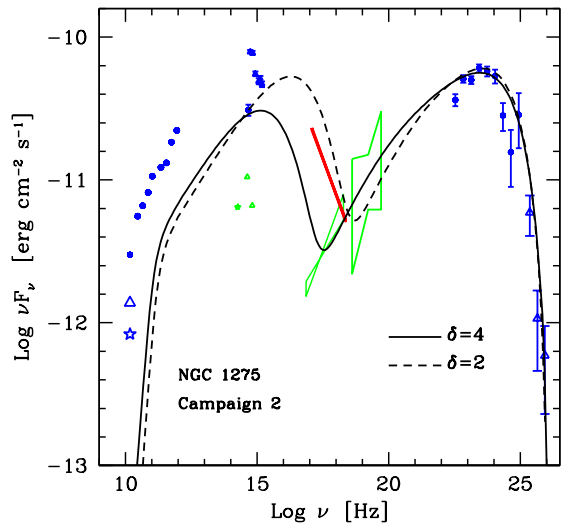}
\caption{NGC\,1275 SED for the epoch of Camp.\,2 (August 2010 -- February 2011). Symbols as in Fig. \ref{fig:SED5}. 
Blue symbols show nearly simultaneous data from MOJAVE, KVA, \textit{Fermi}-LAT and MAGIC analyzed here, and contemporaneous \textit{Swift}-UVOT and \textit{Planck} data from \cite{gio12}. In radio, the level of the sub-pc components C1 (blue star) and C3 (blue triangle) are also shown.}
\label{fig:SED6}
\end{figure}

Nearly simultaneous SEDs for Camp.\,1 and Camp.\,2, assembled with the data described in the previous section, are reported in Fig.~\ref{fig:SED5} (red symbols) and Fig.~\ref{fig:SED6} (blue symbols), respectively. The non-thermal X-ray component spectra from \cite{bal06} (\textit{Chandra}) and \cite{aje09} (\textit{Swift}/BAT), as well as the optical-IR measurement by \textit{HST} \citep{chiab99,baldi}, are also reported (green symbols).

The data hint at a double-peaked SED, similar to the other TeV-emitting AGNs. The similarity of this SED with that of blazars is a clear indication that the broadband emission arises in the relativistic jet. The available data provide an excellent description of the high-energy bump of the SED. In particular, it is possible to constrain the position of the peak with good accuracy.
On the other hand, the low-energy bump is poorly constrained by our radio-to-UV simultaneous data, which are more likely dominated by larger scale emission regions. The correlation between the optical and the $\gamma$-ray LC suggests that the non-thermal continuum from the NGC\,1275 nucleus is dominated by the emission from the same region. We assume that the low-energy component of this region is at the level of our KVA measurements corrected for the host-galaxy and line emissions.

For our model, we assumed, as a starting point, that the emission is coming from a unique uniform spherical region (a bulk plasma) moving along the jet.
We reproduced the SED using a synchrotron self-Compton (SSC) model (for details see \citealt{mt03}) considering an electron-density energy distribution of the form $N(\gamma)=K\gamma ^{-n} \exp(-\gamma/\gamma _{\rm c})$, with a Lorentz factor $\gamma>\gamma _{\rm min}$. Such a very simple distribution is naturally expected in the case of Fermi-I type shock acceleration (e.g. \citealt{b&e87}) and it is also supported by the $\gamma$-ray SED fits of Section~\ref{subsec:3.3}. The other physical parameters specifying the model are the intensity of the magnetic field, $B$ (assumed to be tangled), the source radius, $R$, and the Doppler factor, $\delta$. The latter parameter is given by $\delta=1/[\Gamma_b(1-\beta_b \cos \theta_{\rm v})]$, where $\beta_b$ is bulk speed of the plasma, $\Gamma_b$ is the bulk Lorentz factor, $\Gamma_b = [1-\beta_b^2]^{-1/2}$, and $\theta _{\rm v}$ is the angle between the bulk velocity and the line of sight.
Our model accordingly had only seven independent parameters ($K$, $n$, $\gamma _{\rm min}$, $\gamma _{\rm c}$, $B$, $R$, and $\delta$).

\begin{table*}
 \centering 
       \begin{tabular}{lccccccc}
         \hline
         \hline
           & $\delta$ & $B$ & $K$ & $R$ & $\gamma _{\rm min}$  & $\gamma_{\rm c}$ & $n$ \\
           & & [mG]  &  [10$^4$ cm$^{-3}$] & [10$^{16}$ cm] & & & \\
         \hline
	Camp.\,1 & 4 & 42 & 20 & 7.9 & 100 & $2.2\times 10^5$ & 2.55 \\
                   & 2 & 170 & 18 & 8.2 & 100 & $3.5\times 10^5$ & 2.55 \\
          \hline
	Camp.\,2 & 4 & 38 & 9 & 8.2 & 100 & $0.9\times 10^5$ & 2.4 \\
                   & 2 & 380 & 14.7 & 4.1 & 100 & $1.5\times 10^5$ & 2.4 \\
          \hline
          \hline
         \end{tabular}
         \caption{Parameters for the models reported in Figs.~\ref{fig:SED5}-\ref{fig:SED6}. For both epochs the table reports the result obtained for two different choices of Doppler factor, $\delta=2$ and 4. The parameters are the magnetic field $B$, the particle density $K$, the source radius $R$, the minimum electron Lorentz factor, $\gamma _{\rm min}$, the cutoff Lorentz factor $\gamma _{\rm c}$, and the electron spectrum slope $n$. }
         \label{table:param}
         \end{table*}
         
Following the radio counter-jet evidence \citep{ver94,wal94,asa06}, we first assumed an observing angle $\theta _{\rm v}=30^\circ$, which is the lowest angle compatible with the radio observations. The highest value of the Doppler factor allowed was then $\delta=2$, reached for a bulk Lorentz factor $\Gamma_b=2$. After fixing the Doppler factor, the source size can be constrained by the observed variability timescale in the \textit{Fermi}-LAT band, $t_{\rm var}\simeq 1$ week, which limits the radius through the causality relation $R<c t_{\rm var} \delta\simeq 4 \times 10^{16}$\,cm, rather similar to the sizes used in most blazar models. During Camp.\,1 the \textit{Fermi}-LAT LC is smoother and the constraint on the radius must be relaxed to light-month scales: $R/\delta< 10^{17}$\,cm.

Some estimates of the other physical parameters can be derived with simple arguments. While the peak of the SSC bump arises from scattering in the Klein-Nishina (KN) regime, its frequency can be directly used to estimate the Lorentz factor $\gamma_{\rm c}$ of the electrons at the high-energy cutoff. Following \cite{tmg98}\footnote{\cite{tmg98} considered an electron energy distribution with a broken power-law shape. The results can be extended to the present case of a power law with exponential cutoff by identifying the break parameter $\gamma _b$ with the cutoff parameter $\gamma _c$ and taking the limit $\alpha _1 = \alpha$ and $\alpha _2\gg 1$.}, the frequency of the SSC peak, $\nu _{\rm SSC}$, is given as $h \nu _{\rm SSC} \simeq m_e c^2 \, \gamma _{\rm c} \, g(\alpha) \delta$, where $g(\alpha)=\exp[1/(\alpha-1)]$ is a function of the energy-flux spectral slope of the rising edge of the synchrotron and inverse Compton bumps, $\alpha=(n-1)/2$. The fit of the SSC bump with a power-law with a sub-exponential 
cutoff, reported in Section~\ref{subsec:3.3}, provides a good estimation of the rising edge slope $\alpha\simeq0.75\pm0.1$ (the fit parameter $\Gamma$ is the power-law index of the photon flux, $\Gamma =\alpha+1$). This value also agrees well with the spectral slope measured in hard X-rays with \textit{Swift}-BAT \citep[$\alpha=0.7^{+0.3}_{-0.7}$,][]{aje09}. For $\alpha=0.75$, the value of $g\simeq 0.02$. The SSC peak is measured around $\nu _{\rm SSC}\simeq 10^{24}$\,Hz, implying $\gamma _{c}\simeq 2\times 10^5$. The frequency of the synchrotron peak $\nu _{\rm s}$ is now found as $\nu _{\rm s}\simeq 2.8\times 10^6 B \gamma ^2_{\rm c} \delta$\,Hz. The condition \citep{tmg98} that the SSC peak is produced by scatterings in the KN regime is $h\gamma _{\rm c} \nu_{\rm s}/\delta > m_e c^2$, which, after some operations, can be expressed as a limit for the magnetic field, $B>B_{\rm cr}\gamma _{\rm c}^{-3}$\,G, where $B_{\rm cr}=4.4\times 10^{13}$\,G is the critical magnetic field. Inserting the value above for $\
gamma _{\rm c}$, we found $B>4\times 10^{-3}$\,G.

An estimate of the magnetic field can be derived from the ratio between the synchrotron and SSC luminosities, $L_{\rm s}/L_{\rm SSC}\simeq U_{B}/U_{\rm rad}$, where $U_B=B^2/8\pi$ is the magnetic energy density and $U_{\rm rad}$ is the synchrotron radiation energy density \citep{tmg98}. The latter can be calculated from $U_{\rm rad}\simeq L_{\rm s}/(4\pi R^2 c \delta^4)$. Assuming the SSC emission occurs in the KN regime reduces the radiation energy density available for the emission. However, in first approximation one can neglect this effect (this derived value is formally a lower limit of the magnetic field). Finally, the magnetic field can be expressed as $B\simeq (2/c)^{1/2} R^{-1} \delta ^{-2} L_{\rm s}\, L_{\rm SSC} ^{-1/2}$. Deriving the luminosities from the SED and using $R=4\times 10^{16}$\,cm, we obtain $B\simeq0.1$\,G. This value is well above the limit for the KN regime calculated above, confirming the previous assumption.

Table \ref{table:param} reports the parameters obtained for the emission model, shown as dashed curves in Figs. \ref{fig:SED5}-\ref{fig:SED6}. The frequency of the synchrotron peak is relatively high ($>$10$^{16}$\,Hz) and the model predicts soft X-ray emission dominated by the end of the synchrotron hump. The expected X-ray flux is much higher than the archival \textit{Chandra} data and shows a very different spectral shape.
This would rule out our model if the data were simultaneous.
However, this \textit{Chandra} observation (ObsID 3404) was taken in 2002, long before the renewed activity in radio band started in 2005, and the X-ray emission may have changed dramatically. Compared with the contemporaneous \textit{Chandra} data analyzed in section~\ref{subsec:2.3}, our SED model for Camp.\,1 agrees well when one assumes (as must be done because the strong pileup does not allow us to estimate it) a photon index $\Gamma$=2.5 as expected from the model.
For Camp.\,2, the expected X-ray flux is even higher and the UV emission almost reaches the \textit{Swift}-UVOT data. The simultaneous data do not rule out our model but the overall shape with a synchrotron peak in X-ray seems unlikely. The X-ray flux has never been measured at such a high level. A lower synchrotron peak frequency (in optical) would be more natural. 
It is very important to know the slope of the X-ray spectrum to determine the position of the synchrotron peak. Unfortunately, the quality of the contemporaneous \textit{Chandra} data, which are strongly affected by pileup, does not allow us to derive strong constraints. An X-ray emission from either synchrotron ($\Gamma$=$\sim$2.5) or inverse Compton ($\Gamma$=$\sim$1.7) is possible.

As discussed in \cite{tg08} and \cite{ale10b}, a small ``distance" between the SED peaks is unavoidable for low Doppler factors. In order to increase the separation between the peaks, which would shift the synchrotron bump to lower frequencies, we are forced to increase $\delta$. In fact, using the relations developed above, it is relatively easy to see that (for a fixed $\nu_{\rm SSC}$) the synchrotron peak frequency is related to the corresponding Doppler factor by $\nu_{\rm s}\propto \delta ^{-4}$. In Figs. \ref{fig:SED5}-\ref{fig:SED6} and  Table~\ref{table:param}, we also report the models for a Doppler factor that is twice as large, $\delta=4$ (solid lines). According to the scaling above, the synchrotron peak frequency decreases by more than a decade. The optical flux remains at the same level, while the UV and X-ray emission decrease significantly.Our model with $\delta$=4 would be almost compatible with the hard X-ray spectrum reported in \cite{bal06}. 
This value of the Doppler factor is disfavored, however, since it requires an angle smaller than $15^\circ$ between the plasma bulk velocity and the line of sight.

\section{Discussion}
\label{sec:4}
The overall SED of NGC\,1275 suggests blazar-like emission. A one-zone SSC model can fit the data assuming a jet-viewing angle ($\theta_{\rm v}$\,=\,30$^\circ$) compatible with the radio observation. All parameters (possibly with the exception of the bulk Lorentz factor $\Gamma_b$) derived from this modeling are in the range of the values typically obtained for VHE $\gamma$-ray emitting BL Lac objects (e.g., \citealt{dawn}).
Another interesting remark concerns the required slope of the electron energy distribution, which is close to 2.5. This value agrees well with recent simulations by \citet{sironi11}, who showed that for low values of the jet magnetization (such as those derived by our SED modeling) Fermi-I shock acceleration proceeds efficiently, leading to power-law tails with slope $n=2.5$.

All these elements support the view that the $\gamma$-ray emission of NGC\,1275 originates from a misaligned BL Lac jet, in the same way as for the other VHE $\gamma$-ray emitting radio galaxies, and consistent with the unification scheme (e.g., \citealt{urry&padovani}).
A notable exception, however, is the bulk Lorentz factor, constrained to have a lower value ($\Gamma_b \simeq 2$) than the typical $\Gamma_b>$\,10 inferred in BL Lacs. Indeed, with a jet-viewing angle $\theta_{\rm v}\simeq30^\circ$, a larger $\Gamma_b$ would imply a smaller Doppler factor, while our model with $\delta=2$ already seems at the lower edge of the acceptable range
(a lower $\delta$ would imply a smaller distance between the two SED peaks, inducing a too high X-ray flux).
Such a low $\Gamma_b$ would indicate either that the jet of NGC\,1275 is ejected at a lower speed than in blazars, or that the jet suffers strong acceleration/deceleration in the vicinity of the central engine. 
The kinematics of the radio knots might suggest acceleration along the NGC\,1275 jet \citep{dha98}. Its $\gamma$-ray emitting region might then be closer to the central black hole than for typical blazars (before the bulk is accelerated to its full speed).

However, a larger bulk Lorentz factor would be possible with a smaller jet-sight angle $\theta_{\rm v}$.
For example, a bulk emitting region with $\Gamma_b=10$ would be seen with a Doppler factor $\delta=2$ for $\theta_{\rm v}=17^\circ$ and with $\delta=4$ for $\theta_{\rm v}=12^\circ$.
In the past, a smaller viewing angle has been suggested for the innermost jet \citep{kri92, ver78}. The NGC\,1275 core could then be a BL Lac with strong jet bending at larger scale.
The high energy non-thermal continuum of NGC\,1275 could be also dominated by a sub-structure of the jet that points closer to our direction than the overall radio jet. Such scenarios have previously been proposed to explain the $\gamma$-ray emission from other TeV radio galaxies (\object{M\,87} and \object{Cen\,A}). This sub-structure could originate from the jet formation zone near the central black-hole where the jet is less collimated \citep{len08}, or from mini-jets in the main jet induced by magnetic reconnections \citep{gia10,cui12}. Alternatively, the emission can arise in a magneto-centrifugally accelerated flow (e.g., \citealt{rie02}), whose weak collimation would be compatible with the large viewing angles inferred for radio galaxies.


One may notice that our model predicts a much lower flux at 15\,GHz than the VLBA measurements, even in a single sub-structure. This is related to the relatively high energy peak of the synchrotron component. The radio emission could come from another, more external, region where the low-energy electrons could diffuse. The variability of such an external component should be much smoother.
The similar long-term trend of the C3 component and the GeV $\gamma$-ray emission makes it a suitable candidate for hosting the $\gamma$-ray emitting region.
However, the apparent velocity of this component, $\beta_{app}$=0.23 \citep{nag10}, is much too slow compared with the apparent velocity of the emitting region of our one-zone SSC model. For $\theta _{\rm v}=30^\circ$ and $\Gamma_b=2$, the expected apparent velocity is superluminal,
$\beta_{app}=\frac{\beta_b sin(\theta _{\rm v})}{1-\beta_b cos(\theta _{\rm v})}\simeq1.7$. The models with smaller viewing angle would have even larger $\beta_{app}$. 
Indeed, the C3 radio component and the $\gamma$-ray source may not be co-spatial but the increasing emission in both may simply be connected to a general increasing activity of the AGN. Moreover, no clear correlation on short timescales was found by \cite{nag12}.

The difficulty in determining the correct bulk Lorentz factor of the jet is not specific to NGC\,1275 but is a general problem of the AGN unification scheme, the so-called bulk Lorentz factor crisis \citep{hen06}.
While one-zone SSC models of TeV blazars require a large Lorentz factor ($\Gamma_b>10$), geometrical beaming arguments on FR~I radio galaxy - BL Lac unification predict a low value ($\Gamma_b \simeq 3$) \citep{chiab00}.
More complex models that assume, for example, a structured jet with different emission zones \citep{ghi05}
or blob-speed changes along the jet \citep{geo03}, must be developed to solve this crisis.
The simultaneous broadband observations of NGC\,1275 reported here provide a rare view of a very likely misaligned BL Lac object. They provide new constraints on models developed to reconcile the VHE $\gamma$-ray observations and the unification scheme of AGNs.

\section{Summary and conclusion}
\label{sec:5}
We analyzed multi-frequency observations of NGC\,1275 contemporaneous to two MAGIC observation campaigns
carried out from October 2009 to February 2011.
MAGIC data analysis, with an energy threshold of 100\,GeV, resulted in the detection of the source during both campaigns
separately, confirming that NGC\,1275 belongs to the very exclusive club of radio galaxies detected at VHE.
The $\sim$65-650\,GeV spectra measured with MAGIC are very similar for both campaigns.
The monthly VHE LC shows only a hint of variability (3.6\,$\sigma$) during the first campaign.
\textit{Fermi}-LAT results between 100\,MeV and $\sim$50\,GeV present different features.
The LAT LC shows a clear week-scale variability, and the two derived spectra, corresponding to each of the MAGIC campaigns, show differences in both flux intensity and spectral shape.
Nevertheless, the combined LAT/MAGIC spectra can be well fit by both a log parabola and a power law with a sub-exponential cutoff.
The SED-peak energies of the log parabolic fits are 2.8$\pm$1.5\,GeV (Camp.\,1) and 1.1$\pm$0.2\,GeV (Camp.\,2).
The sub-exponential cutoff energies are 13.1$\pm$4.2\,GeV (Camp.\,1) and 5.0$\pm$1.6\,GeV (Camp.\,2).
The rapid variability and spectral evolution imply that the $\gamma$-ray emitting region is a compact zone in the AGN.
The KVA R-band LC, corrected for host-galaxy and emission-line contaminations, shows a shape very similar to the \textit{Fermi}-LAT LC. This almost linear correlation strongly suggests that the optical counterpart of the very compact $\gamma$-ray emitting region dominates the corrected KVA result.
The 15\,GHz MOJAVE data allowed us to derive the LC of the three innermost radio components. The recently born C3 component shows a similar LC trend as the KVA and LAT LCs. This makes this jet feature a possible candidate for the $\gamma$-ray emitting region.

We modeled the simultaneous broadband SED with a simple one-zone SSC emission model assuming a power-law with an exponential cutoff for the electron energy distribution.
While the high-energy peak is well constrained by the LAT and MAGIC data, at low energy most of our simultaneous data must be considered as upper limits due to possible contaminations. We assumed as optical flux the corrected KVA data, which show correlation with the $\gamma$-ray flux.
The relatively large angle between the NGC\,1275 jet and the line of sight ($\theta_{\rm v}=30^\circ$--$55^\circ$) strongly limits the Doppler factor $\delta$. A model with $\delta=2$ ($\theta_{\rm v}=30^\circ$ and $\Gamma_b=2$) can fit the SED of both epochs and explain the multi-frequency behavior (optical-GeV correlation and rapid variability). However, the expected X-ray level is in contrast with the \textit{Chandra} measurements and models with lower $\delta$ would hardly match the data as they would imply a too small spread between the low- and high-energy SED peaks.
On the other hand, models with larger $\delta$ (such as $\delta=4$) would fit the SED well, but would require smaller $\theta_{\rm v}$.
In fact, the parameters of our models are in the typical range found for BL Lacs, except for the bulk Lorentz factor.
NGC\,1275 might accordingly be a misaligned BL Lac with a particularly low bulk Lorentz factor, or it might be more aligned than we think.
Assuming a smaller $\theta_{\rm v}$ (10--15$^\circ$), larger bulk Lorentz factors ($\Gamma_b>10$) are possible.
A $\gamma$-ray emitting region with a velocity more aligned to the line of sight than the pc-scale radio jet
could be explained by jet bending in the innermost region, a larger jet opening angle near the central black hole, or mini-jets within the main jet.
Multi-zone models would certainly be more appropriate for interpreting the NGC\,1275 AGN emission and its connection to BL Lac objects.

\begin{acknowledgements}
The MAGIC collaboration would like to thank the Instituto de Astrof\'{\i}sica de
Canarias for the excellent working conditions at the
Observatorio del Roque de los Muchachos in La Palma.
The support of the German BMBF and MPG, the Italian INFN, 
the Swiss National Fund SNF, and the Spanish MICINN is 
gratefully acknowledged. This work was also supported by the CPAN CSD2007-00042 and MultiDark
CSD2009-00064 projects of the Spanish Consolider-Ingenio 2010
programme, by grant DO02-353 of the Bulgarian NSF, by grant 127740 of 
the Academy of Finland,
by the DFG Cluster of Excellence ``Origin and Structure of the 
Universe'', by the DFG Collaborative Research Centers SFB823/C4 and SFB876/C3,
and by the Polish MNiSzW grant 745/N-HESS-MAGIC/2010/0.

The \textit{Fermi}-LAT Collaboration acknowledges ongoing
support from a number of agencies and institutes that have
supported both the development and the operation of the LAT as
well as scientific data analysis. These include the National Aeronautics
and Space Administration and the Department of Energy
in the United States, the Commissariat \`a l'Energie Atomique and
the Centre National de la Recherche Scientifique/Institut National
de Physique Nucl\'eaire et de Physique des Particules in France,
the Agenzia Spaziale Italiana and the Istituto Nazionale
di Fisica Nucleare in Italy, the Ministry of Education, Culture,
Sports, Science and Technology (MEXT), High Energy Accelerator
Research Organization (KEK) and Japan Aerospace Exploration
Agency (JAXA) in Japan, and the K. A. Wallenberg
Foundation, the Swedish Research Council, and the Swedish
National Space Board in Sweden.

Additional support for science analysis during the operations
phase is gratefully acknowledged from the Istituto Nazionale di
Astrofisica in Italy and the Centre National d'Etudes Spatiales
in France.

This research has made use of data from the MOJAVE database that is maintained by the MOJAVE team \citep{lis09}.

\end{acknowledgements}

\end{document}